# Digital Transformation: Environmental Friend or Foe? Panel Discussion at the Australasian Conference on Information Systems 2019

**Sachithra Lokuge**

College of Business and Law

RMIT University, Australia

**Vanessa Cooper**

College of Business and Law

RMIT University, Australia

**Darshana Sedera**

Digital Enterprise Lab

Southern Cross University, Australia

**Frada Burstein**

Faculty of Information technology

Monash University, Australia

**Abstract**

The advent of digital technologies such as social media, mobile, analytics, cloud computing and internet-of-things has provided unique opportunities for organizations to engage in innovations that are affordable, easy-to-use, easy-to-learn and easy-to-implement. Transformations through such technologies often have positive impacts on business processes, products and services. As such, organizations have managed to increase productivity and efficiency, reduce cycle time and make substantial gains through digital transformation. Such transformations have also been positively associated with reducing harmful environmental impacts by providing organizations alternative ways of undertaking their business activities. However, in recent times, especially with an abundance of technologies being available at near-zero costs, questions regarding the potential negative impacts of digital transformation on the environment have arisen. The morass of the ubiquitous technologies around us necessitates the continuing creation of large data centers, that are increasing their capacity yielding a negative impact on the environment. Considering this dialectical contradiction, a panel was conducted at the Australasian Conference on Information Systems (ACIS) in Perth, Australia, in 2019. Its aim was to invigorate the dialogue regarding the impact of digital transformation on environmental sustainability and suggested some directions for future research in this area.

# 1 Introduction

*"Humankind has not woven the web of life. We are but one thread within it. Whatever we do to the web, we do to ourselves. All things are bound together. All things connect."* – Chief Seattle

Over the past three decades, there has been a growing awareness of environmental sustainability among organizations (Hanelt et al. 2016). According to the Global Risk Report, climate change has featured prominently for the past five years, highlighting the risk that it has on individuals and on the planet (World Economic Forum 2016). According to World Bank reports, without urgent actions to reduce environmental pollution, climate change could push an additional 100 million people into poverty by 2030 (The World Bank 2019). The proponents of climate change seek stronger legislation and government intervention to deter pollutant organizations and countries. Moreover, societal pressures have forced organizations to introduce corporate social responsibility strategies that facilitate environmental sustainability (Rush et al. 2015). However, environmental sustainability initiatives often fail due to lack of stakeholder awareness, lack of employee participation, lack of accountability in the process, the inability to integrate performance outcomes, complexity of the process and the difficulty in initiating and managing such initiatives (Sedera et al. 2017).

Technology plays an important role in initiating and managing environmental sustainability. On one side, the advent of social media, mobile, analytics, cloud computing and internet-of-things has provided unique opportunities for organizations to engage in environmental sustainability initiatives that are affordable, easy-to-use, easy-to-learn and easy-to-implement (El-Kassar and Singh 2019; Sedera and Lokuge 2017). Transformations through technologies often have positive impacts on business processes, products and services (Lokuge et al. 2019; Majchrzak et al. 2016; Sedera et al. 2016). Further, in terms of environmental sustainability these technologies assist in obtaining accurate and actionable data through internet-of-things and sensors, creating awareness and seeking collaborations through social media, developing better prediction models through business intelligence and deploying solutions through affordable mobile solutions. Indirectly, digital transformation initiatives assist environmental sustainability through supporting better logistics and supply chain management solutions that reduce carbon footprints, better waste management solutions and minimizing manufacturing requirements through three-dimensional printing. Over time, advancements in digital technologies seem to have softened the burden of balancing economic gains and environmental sustainability (Sui and Rejeski 2002). However, with the abundant availability of technologies at near-zero costs, questions have arisen on the potential negative impacts of digital transformation on the environment (Bieser and Hilty 2018). The morass of the ubiquitous technologies around us necessitate the creation of large data centers that are increasing in terms of their capacity and negative impact on the environment. For example, it is predicted that data-center electricity use is likely to increase about 15-fold by 2030, to 8% of projected global demand for electricity (Andrae and Edler 2015). Further, initiatives like bitcoin are said to cause substantial increases in the energy use (Jones 2018).

Considering this debate surrounding the technology's impact on the environment, this panel set out to stimulate the dialogue regarding the impact of digital transformation on environmental sustainability. Despite a wealth of literature on digital transformation and environmental sustainability, much less attention has been devoted to the understanding of the impact of digital transformation on environmental sustainability. A discussion on the theoretical, conceptual and practical notions of environmental sustainability and digital transformation is necessary for researchers as well as practitioners. The panel provided future directions in managing and achieving environmental sustainability goals in digital transformation initiatives. This article presents a summary of an interactive panel conducted at the Australasian Conference on Information Systems, held in Perth, Australia in December 2019. The panel was chaired and moderated by Professor Darshana Sedera of Southern Cross University, Australia and the following members took part as the panelists: Professor Frada Burstein of Monash University, Australia, Professor Vanessa Cooper and Dr Sachithra Lokuge of RMIT University, Australia.

The remainder of this article proceeds as follows. First, an overview of the importance of sustainability in digital transformation is provided, summarizing the discussion of Prof Darshana Sedera. Dr Sachithra Lokuge's discussion that focused on aligning environmental sustainability in strategic digital transformation initiatives is discussed next. Then, Prof Vanessa Cooper's discussion on the capabilities required for environmentally sustainable digital transformation initiatives is outlined. The next section provides the insights into incorporating environmental sustainability in the decision-making process which were discussed by Prof Frada Burstein. Finally, this article concludes by providing a framework for environmentally sustainable digital transformation initiatives and future research ideas for environmental sustainability and digital transformation.

## 2 Why Environmental Sustainability is Important in Digital Transformation? An Overview

The advent of digital technologies such as social media, mobile, analytics, cloud and internet-of-things has enabled digital transformation, a phenomenon that has received attention of academics (Bieser and Hilty 2018; Li et al. 2018; Lokuge and Sedera 2016; Sedera and Lokuge 2019; Vial 2019) as well as practitioners (Forbes Insight 2016; Haffke et al. 2016). Vial (2019, p. 118) defines digital transformation as "a process that aims to improve an entity by triggering significant changes to its properties through combinations of information, computing, communication, and connectivity technologies." As per Wessel et al. (2020), digital transformation is different to an IT strategic initiative as in digital transformations, digital technology plays a central role in redefining value propositions, which triggers the emergence of new organizational identity. The transformed identity of the organization provides positive changes including enhanced decision-making capabilities (Brynjolfsson 2011; Huber 1990), redefined value propositions (Wessel et al. 2020), increased customer connectedness (Bharadwaj et al. 2013; Kumar et al. 2010), expanded channels for reaching customers/suppliers (Bharadwaj 2000; Kleis et al. 2012) and enhanced communication facilities (Olesen and Myers 1999; Youmans and York 2012). While such changes are positive, Sedera highlighted that, despite the advantages, the use of digital technologies in an organization is often associated with negative impacts which are not always taken into consideration. While most researchers extol the positive role of digital technologies, the increased carbon footprint, increased wastage and the damage they cause to the environment are also topics of concern among researchers. Such discourses highlight the urgent need to consider environmental sustainability in digital transformation initiatives.

Sustainability is defined as "development that meets the needs of the present without compromising the ability of future generations to meet their needs" (World Commission on Environment Development 1987). It is vital to have a discussion on environmental sustainability in an era where digitalization has become the number one priority of contemporary organizations. Hence, while initiating digital transformation projects, organizations are required to incorporate environmental sustainability aspects. As a result, there is an ongoing discussion among the scholars in highlighting the importance of environmental sustainability in strategic initiatives for digitalization. The panel acknowledges the two relationships between environmental sustainability and digital transformation: (i) Environmental sustainability through IT and (ii) Environmentally sustainable IT.

Environmental sustainability through IT focuses more on making production processes greener. Here, the focus is on applying more environmentally sustainable practices using IT. For example, introduction of software to measure employees' carbon emission is a novel approach. Such IT initiatives have made employees mindful about their role in achieving environmentally sustainable work practices. On the other hand, environmentally sustainable IT focuses on making IT itself greener. For example, environmentally sustainable IT focuses on green data centers, reduction of greenhouse gas emissions etc. The focus of this panel is on 'environmental sustainability through digital transformation' as this phenomenon is applicable and relevant to all organizations despite their size, industry sector and resourcefulness.

Adhering to environmental sustainability and green management practices are necessary and sometimes mandated to organizations through government initiatives. For example, regulations such as Australian Environment Protection and Biodiversity Conservation Act 1999 provides a legal framework to protect and manage all matters related to national environmental significance. Many organizations perceive environmental sustainability and green management as a 'responsibility' or as a 'compliance' issue rather than seeing it as an opportunity. As such, most organizations view a typical environmental sustainability management initiative as a cost. Therefore, such initiatives fail to gain traction with key stakeholders and wither without achieving the proposed environmental effects. Moreover, the 'cost' perspective fails to make such initiatives valuable to the organization. However, several real-world examples highlight the importance and the value of following environmentally sustainable practices in an organization. For example, Good Guys Capalaba (in Australia) – a less known franchised whitegoods store – initiated an in-store polystyrene recycling program that recycles polystyrene waste to make coat hangers and picture frames, which reduced their carbon footprint tremendously. Moreover, this initiative has reduced approximately 5 tons of waste from Australian landfill annually (https://www.ehp.qld.gov.au).

Research suggests that sustainability initiatives present an opportunity for an organization to think outside the box (Du et al. 2007; Hong et al. 2010; Tushi et al. 2014). There is evidence, for example in the airline and tourism sectors, that customers are willing to pay an extra premium for products or services that are labeled as sustainable. Moreover, there are substantial initiatives by governments to provide incentives for sustainability programs. A recent study observes the interplay between the policies of local and central governments with the behavior of polluting firms and third-party enterprises (Al-Saleh and Mahroum 2015). Moreover, studies also have reported that the focus on environmental sustainability has led to reductions in the long-term cost and gaining cost efficiencies (Ambec and Lanoie 2008). As per Suchman (1995), organizations that are desirable and conduct rightful business certainly receive the support from the external entities and have better access to resources. Organizations that proactively follow environmentally sustainable practices are more likely to gain external support from governments, non-government organizations and general public as they prioritize the environmental concerns (Luo and Du 2012).

Sedera commenced the panel by providing an overarching approach through the Henderson and Venkatraman (1993) IT-business strategic alignment model (SAM). Therein, he argued that, for both academics and practitioners, 'environmental sustainability' should not be an afterthought or an obligation, rather a central component that should be embedded in one's strategy discussion. He argued that the forcible or retrofitting inclusion of environmental sustainability to the strategy discussion – instead of considering environmental sustainability as a component of strategy derivation process – will not change the view of seeing sustainability as a 'cost' or as a 'reduction in profits.'

Proposing the basis of the panel, using the SAM, Sedera suggested that environmental sustainability could be the deciding factor in business strategy, IT strategy, business and IT processes leading to competitive advantage. Though environmental sustainability is a key driver of the contemporary business landscape, the SAM fails to capture the importance of environmental sustainability. Prior researchers have investigated IT alignment from multiple perspectives such as, alignment between business strategy and IT strategy (Chan et al. 2006), business strategy and IT capabilities (McLaren et al. 2011) and IT business alignment of multi-business organizations (Queiroz et al. 2018). However, rarely we see environmental sustainability as a key component considered with the SAM's perspectives of strategic initiatives.

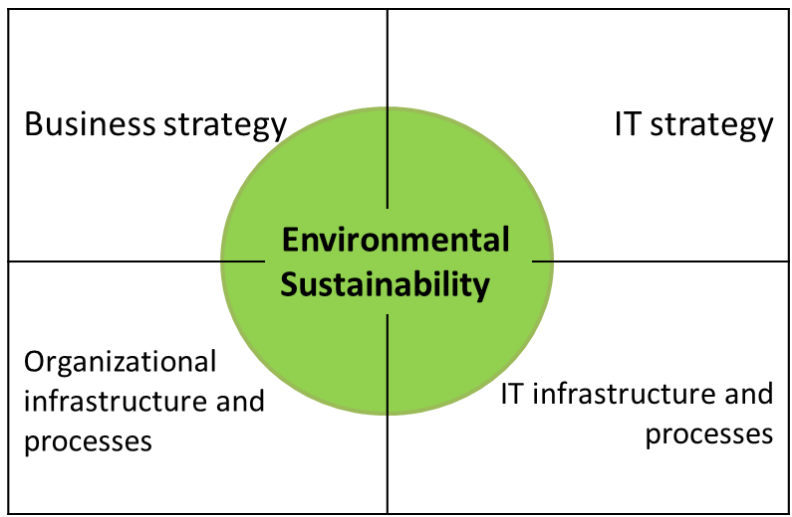

**Figure 1. Proposed Sustainable Strategic Alignment Model**

Keeping environmental sustainability as the core, the remaining panelists contributed to the discussion in the following manner. First, Lokuge extended the notion of the environmental sustainability centrality in the IT-business strategic alignment model. Lokuge in her discussion, proposed an updated model for strategic alignment model with sustainability as the central component. Second, Cooper discussed the capabilities organizations require to ensure environmentally sustainable digital transformations. In so doing, she emphasized the importance of developing capabilities to assess whether a digital transformation will have a positive and/or negative impact on environmental sustainability in the first instance. Third, Burstein discussed the strategic decision-making process in relation to the orchestration of organizational and IT infrastructure and processes to yield an environmentally sustainable practice.

## 3 Alignment of Digital Transformation Initiatives to Environmental Sustainability

Following the central theme suggested by Sedera, Lokuge further explained ways to incorporate environmental sustainability in extending the cross-domain perspectives introduced by Henderson and Venkatraman (1993). In IT-business alignment, the 'alignment' refers to the degree to which the needs, demands, goals, objectives, and/or structures of IT are consistent with the business (Gerow et al. 2014; Gerow et al. 2015). According to Henderson and Venkatraman (1993) organizations should align IT strategy, business strategy, business infrastructure and processes and IT infrastructure and processes to harvest the full potential of their IT strategic initiatives. Further, when commencing digital transformation projects, organizations should ensure they incorporate environmental sustainability into every aspect from planning to execution. So, how do organizations integrate sustainability into SAM? Table 1 below is a summary of what Lokuge suggested as examples where environmental sustainability can be incorporated into strategic alignment model.

Table 1. Examples for enhanced Sustainable Strategic Alignment Model (SAM)

| SAM Component | SAM Sub-component | Examples of relevant environmental sustainability factors |
|---|---|---|
| Business Strategy | Business scope | Sustainability hackathons to identify opportunities and business areas, Reevaluating products and services to incorporate sustainability, Integrated thinking to minimize the impact on the environment, Incorporation of sustainable development goals, Corporate social responsibility goals, Promote green vendors. |
| | Distinctive competencies | Identification of the strengths, weaknesses, opportunities and threats in terms of sustainability, Improving the Green awareness of employees, Introducing sustainability concepts to the competent areas of the business, Introducing a green team to develop and implement sustainable solutions, Eco-branding, Environmental stewardships. |
| | Business governance | Introduction of compliance, governance structures, standards, frameworks and structures to promote sustainability in work practices, Introduction of Greenpeace head for strategic initiatives. |
| IT Strategy | Technology scope | Green hackathons to identify the best technological solutions that incorporates sustainability, Promoting the use of sustainable technologies, Introduction of mandatory guidelines for environment pollution management. |
| | Systematic competencies | Competence in using sustainable IT solutions, Sustainable IS knowledge, Skills Framework for the Information Age. |
| | IT governance | Centralized management of high carbon emitting IT tools, governance structures based on the carbon footprint, Guidelines and framework for new sustainable initiatives, Introduction of carbon emission management plan for technologies. |
| Organizational infrastructure and processes | Administrative infrastructure | Incentivizing individuals for promoting sustainable behaviors, Green IT outsourcing based on carbon emission, Waste management, Optimizing resource usage. |

| | Processes | Greening of all operational activities such as accounting, marketing, supply chain, production etc., Introducing green/sustainability component to performance reviews. |
|---|---|---|
| | Skills | Introduction of training sessions to improve the knowledge of the employees on sustainability practices, promote obtaining certifications for sustainable practices. |
| IT infrastructure and processes | Architecture | Minimizing wastage, use of sustainable IT solutions in the organization, reusing IT, Waste management, Optimizing resource usage, Recycling assets. |
| | Processes | Green business process management, Sustainability concepts in the automation process, Green supply chain management. |
| | Skills | Introducing green challenges to increase awareness of sustainable initiatives, Compliance leadership. |

As per Wessel et al. (2020) digital transformation is a strategic initiative in an organization. Lokuge argued that for digital transformation projects to be successful, the same logic of maintaining IT-business alignment applies. When the components of IT business alignment model are 'aligned' well, it is considered that the organizations are more likely to invest in IT and utilize IT to gain agility, and thereby creating a sustainable competitive advantage (Lokuge and Sedera 2019; Lokuge and Sedera 2020; Sabherwal and Chan 2001). However, even though environmental sustainability is considered as important in such strategic initiatives, this focus is missing. Most organizations are solely focused on the profit, rather than considering the long-term environmental sustainability gains.

Considering the literature on IT business alignment, Lokuge argued the possibility of hypothesizing four dominant alignment perspectives for attaining environmental sustainability. Figure 2 below provides an overview of four alignment perspectives probable for environmental sustainability. The four alignment perspectives have been identified based on the leading component. As per Figure 2 we propose Business strategy and the IT strategy as the catalysts in the determining alignment perspectives.

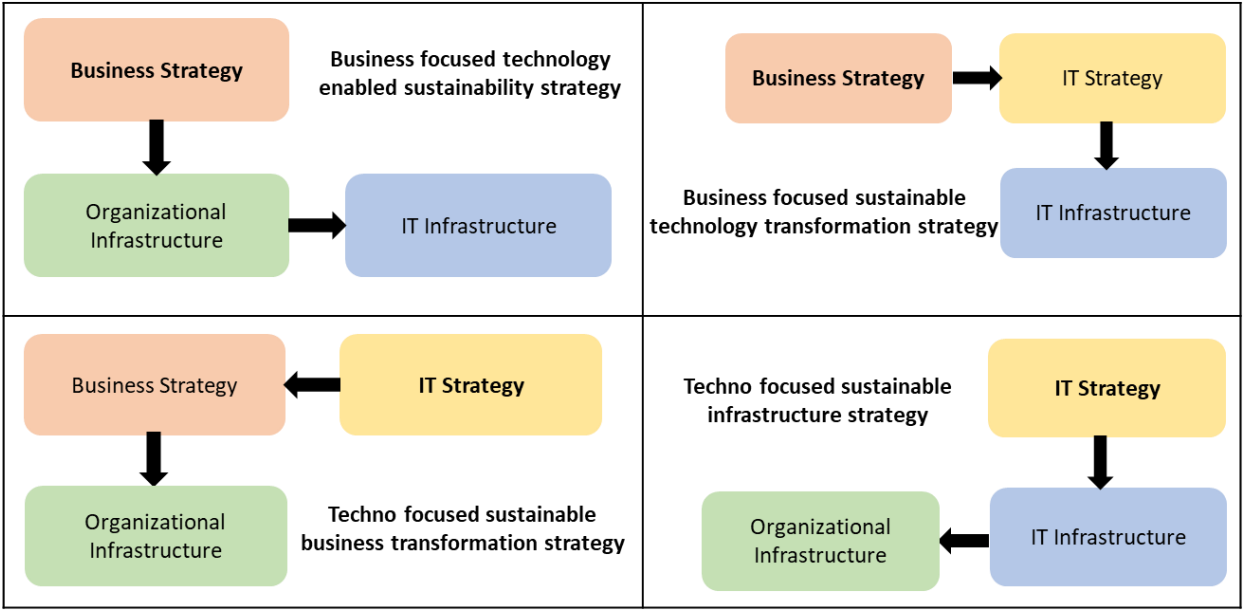

Figure 2. Four alignment perspectives adapted for attaining environmental sustainability

**Business focused technology enabled sustainability strategy**: Environmental sustainability is at the core of the business strategy and it leads the digital transformation execution to be greener. The strategic decision-makers play a key role in leading the environmental sustainability discussion in this perspective. At the executive level, it is important to be aware of the necessity of including sustainability into strategy formation.

**Business focused sustainable technology transformation strategy**: Business strategy maintaining environmental sustainability at the core, driving the digital transformation to be greener. As such, more sustainable IT will be incorporated into the organization as a result of digital transformation. In here, the green business strategy will empower greener IT strategy which ultimately inspire the inclusion of sustainable IT infrastructure in the organization.

**Techno focused sustainable business transformation strategy**: IT strategy is promoting green digital transformation at the organization, thus making the processes, tools and infrastructure of the organization greener. In here, the IT strategy is leading the environmental sustainability discussion in the organization. At the organizational level, inclusion of sustainable IT and thereby the changes in the organizational infrastructure can be predicted.

**Techno focused sustainable infrastructure strategy**: IT strategy drives environmental sustainability discussion in the organization. However, in here this is considered as sustainable IT movement as it makes both IT infrastructure and administration infrastructure greener. As such, it can be argued that the green impact of this strategy is high compared to the other strategies. For this alignment perspective to be successful, global, national, organizational and individual level commitment is important.

## 4 Capabilities for Environmentally Sustainable Digital Transformation

Cooper focused on the importance of organizations developing their capabilities to ensure that digital transformation has a positive impact on environmental sustainability. Picking up on Sedera's introduction, Cooper provided several relatable examples to illustrate the impact of IT on the environment. For example, she pointed to those that may be considered fairly obvious and intuitive to end-users, such as a reduction in the usage of paper and the generation of e-Waste, as well as those that are arguably less visible, such as the amount of $CO_2$ emitted from the use of technology devices (Ansari et al. 2010; Degirmenci and Recker 2018). Cooper emphasized that the impact of IT on the environment is not always straight forward. For example, although users may have the best of intentions in substituting their use of printed reports with digital ones, if this substitution results in users repeatedly downloading reports, then the reduction in paper consumption may be offset by an increase in energy consumption. For example, how often do people check their bank account balances now that they are available online as opposed to when people had to phone or visit the bank? An interrogation of how IT changes behavior and the impacts of these behavioral changes is required because where IT makes behaviors easier to perform, the environment may suffer consequently. In order to determine whether digital transformation is indeed an 'environmental friend or foe' more accurate measures of the environmental impact of IT are required. Cooper illustrated this point using the example of a controversial report in the Sunday Times a decade ago that two Google searches produce the same amount of $CO_2$ as boiling a kettle (a report later scrutinized by Google and the lead researcher cited in the article, Harvard Professor, Dr Alex Wissner-Gross) (Kincaid 2009; Miguel 2009). Examples were used to highlight the complexity of measuring the impact of digital transformation (Bieser and Hilty 2018) and the need for organizations to develop their capabilities to not only measure this impact but to ensure that digital transformations have a positive rather than negative impact on environmental outcomes (Bieser and Hilty 2018; Hanelt et al. 2017).

When focusing on environmentally sustainable digital transformation, the IT capabilities of an organization play an important role. We define IT capability as "the firm's ability to mobilize and deploy its IT-based resources, creating value in combination with other resources and capabilities (Bharadwaj 2000, p. 171), and the firm-specific IT enabled knowledge and routines that improve the value of non-IT resources" (Drnevich and Croson 2013, p. 485). The typology of IS resources (i.e. assets and capabilities) provided by Wade and Hulland (2004) which comprises outside-in, spanning and inside-out areas of capability was used as an example. Although recognizing the value of early work in IT capability, most of the early research in this area was undertaken prior to the Internet-era (Li et al. 2018); and information systems (IS) researchers' interest in "Green IT" and "Green IS." As such, Cooper's presentation turned to a consideration of the capabilities required for digital transformation and for environmental sustainability.

### 4.1 Capability and Digital Transformation

In the digital age, technology is increasingly at the center of how organizations produce value, generate income, and realize competitive advantage. Recent advances in digital technology, platforms and ecosystems (Vial 2019) have extended the breadth and depth of IT's impact on organizations (Lokuge and Sedera 2018). Customers increasingly demand personalized and seamless multi-channel experiences. Rather than simply automating existing business processes, digital transformation changes the digital identity of the organization (Lokuge and Sedera 2014a; Lokuge and Sedera 2014b; Wessel et al. 2020). Under such conditions distinct capabilities are required.

Both researchers and practitioners have reported that today's organizations increasingly require capabilities for developing digital strategy (Lopez 2014), digital customer engagement (Catlin et al. 2015), digital leadership and technology (Lokuge et al. 2018; Lopez 2014), modular IT platforms/platform utilization (Catlin et al. 2015; Li et al. 2018), agile technology-delivery skills (Catlin et al. 2015; Walther et al. 2018), dynamic managerial capabilities (Li et al. 2018), business development capabilities (Li et al. 2018), IT human resource capability and new service delivery capabilities (Aral and Weill 2007; Singh et al. 2011; Walther et al. 2015). Dynamic capabilities, being "the firm's ability to integrate, build and reconfigure internal and external competences to address rapidly changing environments" (Teece et al. 1997) remain central to digital transformation. Cooper highlighted the role of external organizations (e.g., platform providers) in digital transformation, given the complex business eco-systems in which digital transformation takes place (Li et al. 2018). Accordingly, if developing capabilities is important for digital transformation, the question for those interested in environmental sustainability is *what capabilities are required by organizations to ensure digital transformations have a positive rather than negative impact on the environment?*

## 4.2 Capabilities for Environmental Sustainability and Digital Transformation

Despite a growing body of literature on Green IS (Corbett 2013; Watson et al. 2010), with few exceptions (Bose and Luo 2011; Cooper and Molla 2017; Molla et al. 2011), relatively little attention has been given to the specific capabilities required by organizations to leverage IS for environmentally driven digital transformations. For organizations to ensure that digital transformations deliver environmentally sustainable outcomes they must embed environmental sustainability considerations in their IS infrastructure and practices (Hu et al. 2016; Melville 2010) and develop IS innovations that provide environmental benefits (El-Kassar and Singh 2019; Sui and Rejeski 2002; Venable et al. 2011). Developing "Green IS" capability is a specific and complex organizational competence that is different from the development of IS capability for conventional business outcomes. While there is overlap in the processes required to develop capability in Green IS and IS capability in other contexts, IS practitioners should pay careful attention to the differences.

Organizations should not assume that the traditional knowledge and skills of IS professionals are enough to address environmental sustainability challenges, as these are a relatively new concern for IS professionals. IS professionals require "Sustainable IS knowledge" which includes the principles of IS strategy, solutions and evaluation for environmental sustainability (Cooper and Molla 2017). IS educators and professional associations should ensure that environmental sustainability topics are incorporated into IS curricula (Sendall et al. 2011; Watson et al. 2010). For example, as discussed by Lokuge, environmental sustainability introduces a distinctive dimension to the IT–business alignment equation and requires IS professionals and IS departments to extend their traditional knowledgebase and skillsets to develop new capabilities. Frameworks such as the "Skills Framework for the Information Age" offer important guidelines to academics and practitioners. It was not until "SFIA4" that sustainability skills (sustainability strategy, sustainability management, sustainability assessment and sustainability engineering) were introduced into the framework with the most recent revision "SFIA7" seeing these skills merged within more traditional IS skills. While the underlying assumption that sustainability skills are covered in a range of other skill areas may represent an ideal scenario, Cooper argued that this is not without its risks. It may be, for example, that sustainability skills are more easily overlooked. Like many other IS-phenomena, the factors that facilitate and inhibit Green IS capability development need careful attention. There are some unique considerations in the environmental sustainability context. First, unlike market-based resources and capabilities that should be limited to imitation, maximizing the outcome of Green IS capability depends on its diffusion, that is, organizations should share their knowledge and collaborate so that they can collectively address environmental issues (Cooper and Molla 2017). Thus, traditional market forces may produce some nuanced results in this context. Second, where Green IS is viewed as a trade-off with core areas of responsibility (e.g., security, risk management, customer service), Green IS may be deprioritized and thus the facilitators and inhibitors of Green IS capability may have less influence.

To evidence these points, Cooper elaborated on a study she undertook which investigated IS absorptive capacity for environmentally driven IS-enabled transformation (Cooper and Molla 2017). Through a survey of 148 senior IS managers, this study developed a model that explains that IS triggers, knowledge exposure and prior experience influence the development of IS-environmental absorptive capacity, which in turn contributes to the level of environmentally sustainable IS assimilation as well as to the cost saving, operational performance and reputation of organizations. The case study results emphasized the importance of contextual factors at the IS department and organizational levels. For example, at the IS department level a "service provider mindset" was evident where participants viewed the role of the IS department to deliver projects requested at the organizational level rather than to be a thought leader, including those in the sustainability domain. Further, it was found that without clear sustainability commitment, provision of sustainability performance indicators or sustainability championship at the organizational level, the development of IS absorptive capacity for environmentally driven IS-enabled transformation would be hindered. While these findings may be unsurprising, they are indicative of some significant challenges facing those who are serious about ensuring digital transformations are environmentally sustainable and the capabilities that must be developed to overcome them.

The ultimate outcome of leveraging IS for environmental sustainability should be its contribution to the quality of the environment. Returning to the question of whether digital transformation is an environmental friend or foe, it can be noted the causes of environmental outcomes are ambiguous, and deciding what to measure in the first instance is not always straightforward, with differing views across stakeholder groups about what constitutes desirable environmental outcomes and metrics (Cooper and Molla 2017). Further, the complexity of these and other issues require organizations to develop their capabilities in Green IS and it is important that IS researchers contribute further understanding in this area.

## 5 Integrating Environmental Sustainability in IT Decision-Making Process

Burstein contributed to the panel by discussing the opportunities and issues faced in strategizing and decision-making processes in attaining environmental sustainability. Her discussion was based on the fundamental premise of decision-making philosophy of data, decision maker and the decision-making process. The extant literature investigates technical aspects such as integrating lifecycle assessments to costing systems (Tsai et al. 2015), green decision-making models to logistics (Vahabzadeh et al. 2015) and tools for optimizing green building features (Ewing and Baker 2009). Burstein highlighted that as IS researchers, we seldom extend research to enhance capabilities to improve decision-making process regarding attaining environmental sustainability. In addition, the extant research fails to incorporate and investigate the impact of digital technologies and their relevance to strategic decision-making regarding environmental sustainability initiatives. In a time like this, we believe it is important to discuss how organizations could incorporate novel practices that favor green decision-making.

In relation to data, Burstein commenced by reminding the panel participants about the wealth of opportunities that digital technologies such as social media, mobile technologies, analytics and internet-of-things have provided to organizations to receive data of incredible value (Nylén and Holmström 2015). For example, organizations have opportunities to seamlessly gather data about customers, products, business processes and services. Such data can then be processed, to allow decision makers to make effective and informed decisions with an emphasis on sustainability (Lokuge et al. 2020). However, organizations utilize this continuous, rich and voluminous data, to obtain strategizing opportunities but not utilizing them to gain insights for environmental sustainability initiatives. In highlighting that, organizations have the opportunity to look for new pathways to attain their sustainability goals. Therein, the managers (decision makers) must emphasize not only to capture data necessary for financial profitability, but data that concerns environmental perspective as well.

The traditional decision makers now have the added responsibility of being 'environmentally sensible' to take leadership of attaining sustainability goals (Joshi et al. 2003; Kim et al. 2020). Due to constant pressures received from customers (Lieb and Lieb 2010), environmental groups (McKinnon 2010a), public policies (McKinnon 2010b) and global mandates (Turnhout et al. 2016), organizations are compelled to adhere to environmentally sustainable business operations, without compromising the profitability and efficiency of the organization. Traditionally, the decision makers follow approaches such as lifecycle assessment and net present value for assessing the "greenness" of the decision outcomes (Melville and Zik 2016). However, it was noted that such retrospective thinking in decision making rarely favors environmental sustainability. Especially, the middle level and line-of-business managers are less likely to initiate environmental sustainability programs at the expense of compromising efficiency and profitability (Kim et al. 2020). For environmentally sustainable projects to be effective, such directives and support must come from the executive level managers (de Medeiros et al. 2014) incorporating assurances, support, incentives into the organizational policy and procedures (Molla and Abareshi 2012). Once top management support is ensured and a corporate environmental sustainability is entrenched in the policy and procedures, the line-of-business managers could then initiate, fine-tune and manage their sustainability initiatives. The inclusion of clear parameters for looking at relevant sustainability data should be the new "normal" when formulating strategic decision-making.

In line with this, Burstein discussed the need to conduct design science research in integrating environmental sustainability to decision-making process and proposed appropriate decision support systems designs, which include sustainability as one of the design principles. Prior research for example, Seidel et al. (2018) has also proposed design principles for IS that support organizational sensemaking in environmental sustainability transformations. Degirmenci and Recker (2016) have investigated how actual behaviors and decisions of system users can factually be environmentally sustainable through information systems. Further, Melville and Zik (2016) applied design science research to propose an energy productivity approach based on source energy, where they developed a new metric called Energy Points. While these researchers have initiated a discussion surrounding this, finding answers to such problems requires a rethink and convergence of multiple point of views. By applying the systematic research gap analysis approach (Fielt et al. 2014), Burstein discussed how sustainability can be incorporated to decision-making process. For example, from the exploration phase explicitly stating the need and importance to consider environmental sustainability will make the initiatives a success.

Burstein highlighted how environmental sustainability can be incorporated to decision-making, the importance of this process, the impact of it and the critical success factors for such incorporations. She suggested that there is an urgent need to come up with conceptual clarity around relevant classical "What/Why/Who/How" questions. Specifically, such efforts would target:

- *What* – is there a clear definition of a problem space including criteria for digital transformation decision-making, including environmental sustainability as the main objective?
- *Why* – are the reasons, objectives and constraints for digital transformation defined?
- *Who* – is involved in terms of the roles, structure and a flow of authorities for digital transformation decision-making?
- *How* – should decision-making processes (existing and proposed) be adjusted to reflect the concepts above.

The decision-making process is a critical step in digital transformation projects. As discussed above, in the decision-making process the decision-maker, the data/information available for making the decision and the problem that requires attention are important. When introducing and incorporating environmental sustainability into the decision-making process, all three components needs to be considered. For example, the awareness of green initiatives may influence the decision-maker to incorporate green concepts in the decision-making process. While the advent of digital technologies has opened new ways for organizations to collect data, organizations have failed to fully utilize the opportunity of incorporating big data for strategic decision-making. Contemporary research focuses on applying big data concepts to market intelligence, e-governance, health and security areas. In addition, researchers can apply big data to assess greenness of future initiatives. As proposed by Melville and Zik (2016) organizations can collect large data sets on environmental metrics to analyze and derive new metrics related to environmental sustainability. As such, researchers could derive new approaches to comparing different types of energy and sustainability projects that enable better understanding and modeling decision-making situations. Further, when looking at the problem, the decision-maker could consider environmentally sustainable solutions, considering the future possibilities. While all these suggestions may seem like a far cry, it is high time that decision-makers prioritize environmental sustainability as there will be no business without a planet.

# 6 Conclusion

The advent of digital technologies has provided a myriad of avenues for organizations to transform their businesses. As such, in recent times, terms such as digitization, digitalization and digital transformation become buzzwords in both academia and practice. While such terms are associated with organizational performance, efficiencies and productivity, there is a growing concern of the impact of digital transformation on environmental sustainability. In this panel, we discussed the importance of developing an integrated view that aligns sustainability with digital transformation. We acknowledged that the common understanding amongst scholars is that with the growing usage of technologies will inevitably increase the energy consumption, as such the e-wastage and increased carbon footprint (Guster et al. 2009; Sedera et al. 2017). Further, research shows that technology initiatives incur great stress on the environment (Fuchs 2008), calling for researchers and practitioners to look for ways to respond to the growing environmental sustainability issue (Wang et al. 2015). As a result, organizations have experienced considerable global, local and social pressure to initiate environmentally sustainable initiatives to minimize the negative impact of IT on the environment (Nishant et al. 2012). To minimize this pressure, some organizations employ IT to reduce their operational impact on the environment (Hasan et al. 2009) and some organizations utilize environmentally sustainable IT solutions to minimize the impact (Baek and Chilimbi 2010). This panel commenced with the premise that even though researchers have focused on organizational performance aspects of digital transformation, understanding the impact of digital transformation on environmental sustainability has been lacking. This panel inspired a conversation on the impact of digital transformation on environmental sustainability for researchers and practice. A discussion on the theoretical, conceptual and practical notions of environmental sustainability and digital transformation is necessary for the researchers as well as practitioners.

## 6.1 Consolidation Frameworks

The panel first provided an overview of the digital transformation and its impact on environmental sustainability. In here Sedera highlighted the positive and negative effects of digital transformation initiatives on environmental sustainability. Lokuge proposed an extension to IT business alignment model to incorporate environmental sustainability. Then, Cooper discussed capabilities that are required for environmentally sustainable digital transformation initiatives. Finally, Burstein discussed how organizations could incorporate environmental sustainability to decision-making process. In conclusion, the panel highlighted various facets of digital transformation to environmental sustainability, using SAM as the founding theoretical premise. The panel then expanded its views, in four related areas: (i) considering environmental sustainability as the central component of business and IT strategy derivation, (ii) development of IT and business capabilities in alignment with environmental sustainability priorities of the organization, (iii) aligning IT and business capabilities through operationally grounded actions, (iv) decision maker, decision making process and its philosophy embedded in the environmental sustainability priorities and green sensemaking enabled through the digitalization process.

Based on the presentations of the panelists and the comments received from the participants, a framework is derived and depicted in Figure 3. Such a framework would facilitate a holistic understanding of environmentally sustainable action-points from four interrelated levels: individual, organizational, country and global. Note that the central concept of 'Green strategic priorities of the organization' is represented by the 'green strategy formation' of Figure 3.

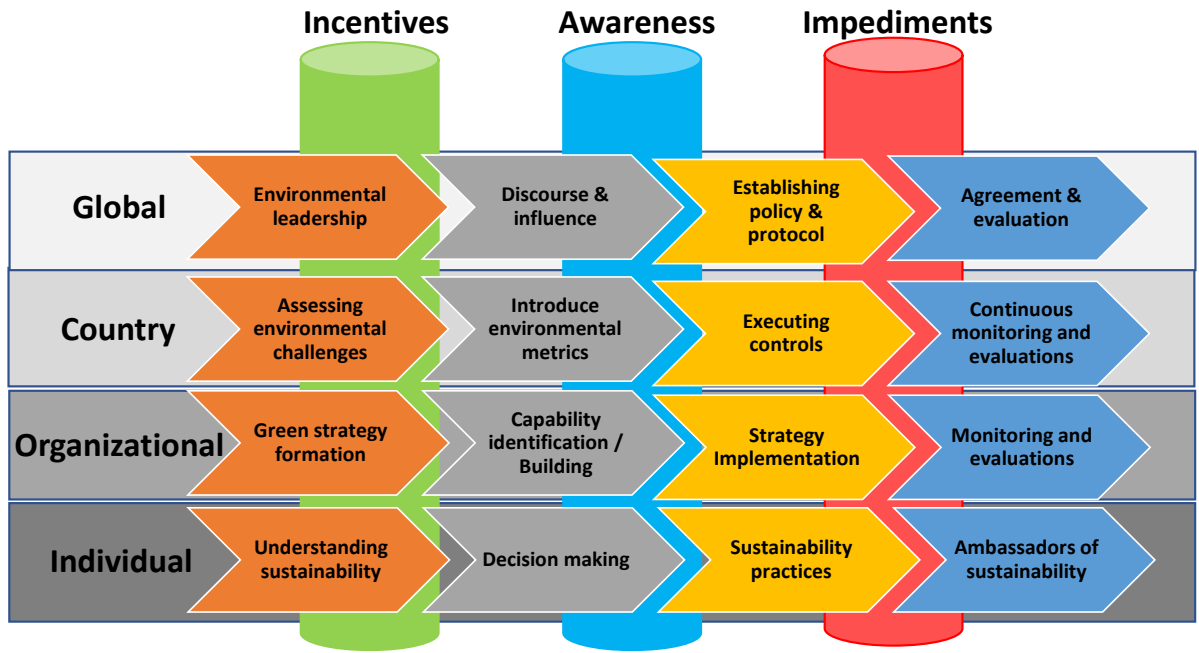

**Figure 3. Consolidated framework for environmentally sustainable digital transformation**

The panel argues that a three-pillar strategy of *awareness*, *incentives* and *impediments* at four levels may be useful. Unlike any other issue or notion, 'sustainability' requires a global coordination that is aligned with the local, organization and individual levels. Awareness refers to the knowledge about the impact of digital technologies and initiatives on environment. The incentive refers to the motivations for conducting sustainable digital transformations. Impediments refers to obstructions for sustainable digital transformations. We propose four processes for obtaining sustainable digital transformations for individual, organizational, country and global levels. While in the panel we focused only individual and organizational level, we extended and proposed the processes for country and global level. While these processes are still in ideation phase, empirical investigations are required to establish these processes.

Digital transformation, whether at the organization or country level has potentially negative impacts on the environment. The increasing footprint of the Internet, heavy use of IT infrastructure and growing digital waste, have the potential to pollute the earth, air and water. However, there is largely an agreement that digital transformation is an essential component of organizations, societies and individuals. As such, the panel argues that we must find a common ground where these two concepts of digital transformation and environmental sustainability can co-exist.

## 6.2 Future Research Areas

The panelists agreed that IS researchers need to pay additional attention to the virtue of digital transformation and environmental sustainability. Prior studies have paid attention to importance of green innovation (Lampikoski et al. 2014; Schiederig et al. 2012), green orientation (Hong et al. 2009), green implementation frameworks (Bose and Luo 2011), organizational support for green management (Loeser et al. 2017) and environmental corporate social responsibility (Ambec and Lanoie 2008). The current topic of digital transformation is timely and of benefit to multiple parties such as individuals, organizations, communities and governments, while Green IS research is matured enough to form multi-disciplinary initiatives to look at the solid science behind the greening efforts, as well as long term implications of such efforts. The proposed framework has been derived through the observations and comments from the panel session. As such, it opens pathways for researchers to contribute to academic knowledge and inform better industry practices.

According to Kappelman et al. (2014), understanding business and business requirements and maintaining the IT capabilities to survive in dynamic business environments is the third most mentioned issue for IT executives. In a highly volatile business environment, with pressure emerging from the external institution to adhere to environmental regulations, organizations are under immense pressure to survive. While adhering to green policies, aligning their strategic objectives requires organizations to develop and leverage their IT capabilities. As such, researchers have the potential to investigate the following question:

*RQ1: What is the role of IS in facilitating environmentally sustainable digital transformation initiatives?*

Answering such a question would involve exploration of several related questions, including:

*RQ2: How does IT facilitate organizations to balance environmentally sustainable IT/business alignment?*

Brendel et al. (2018) analyzed previous research efforts since 2007 which produced design artifacts to address green IT. Their observations on the future opportunities for addressing outstanding research issues in the impact of digital artefacts are highly relevant to the propositions derived by this panel. They noted the lack of theoretical knowledge development and reflections on the implications of the IT on sustainable future. This correlates to our proposition that environmentally sustainable digital transformation research efforts so far were mostly atheoretical. Hence, the researchers could aim to work towards a theory that will extend the green management, strategic literature and IT capabilities body of knowledge. We propose researchers to address this gap by answering the question:

*RQ3: What are relevant theories for describing, explaining, predicting and/or prescribing environmentally sustainable digital transformation practices?*

In doing so there is a need to determine if there are any indigenous, IS-specific theories which are valid for describing, explaining, predicting and/or prescribing environmentally sustainable digital transformation practices, with a particular focus on design theories and, importantly, increasing the rigor and numbers of quality publications in this area. For IS researchers this also provides an opportunity to engage with other relevant disciplines that are required to inform multidisciplinary environmental sustainability agenda in digital transformation context.

Although big data is a common topic of interest, IS researchers have rarely investigated the application of bigdata in the area of environmental sustainability. As such, there is a great opportunity for pioneering research and providing new insights. Further, given the lack of prior research, there is a clear opportunity to contribute to the environmental sustainability research stream by employing a design science approach focused on strategic decision-making using big data. Continuing the Brendel et al (2018) argument about the lack of behavioral research as critical success factor on impact of new IT systems on individuals, there is a clear need for an environmental sustainability design science research focused on strategic decision-making. Using relevant sources of relevant historical (big) data should lead to practice-driven insights on the implications of digital transformation on individuals, organizations, countries, as well as globally, in line with the consolidated framework proposed by our panel. Thus, the research question to address can be:

*RQ4: What factors can influence practical steps in changing human decision-making behavior leading to sustainable digital transformation?*

We call for future research to conduct a systematic literature review including academic, as well as practitioner literature to identify a variety of case studies which describe the ways environmental sustainability issues were covered in digital transformation projects. Sound methodological guidelines and practitioner-focused policies could be created and initiated in technical artefacts as part of research and development efforts for sustainable digital transformation. The other opportunity mentioned was proposing suitable IS curricular to be included in the training of a new generation of environmentally conscious, socially responsible IT professionals for the future.

# References

Al-Saleh, Y., and Mahroum, S. 2015. "A Critical Review of the Interplay between Policy Instruments and Business Models: Greening the Built Environment a Case in Point," *Journal of Cleaner Production* (109), pp 260-270.

Ambec, S., and Lanoie, P. 2008. "Does It Pay to Be Green? A Systematic Overview," *The Academy of Management Perspectives* (22:4), pp 45-62.

Andrae, A.S., and Edler, T. 2015. "On Global Electricity Usage of Communication Technology: Trends to 2030," *Challenges* (6:1), pp 117-157.

Ansari, N.L., Ashraf, M.M., Malik, B.T., and Grunfeld, H. 2010. "Green IT Awareness and Practices: Results from a Field Study on Mobile Phone Related E-Waste in Bangladesh," *IEEE International Symposium on Technology and Society*, Wollongong, NSW, Australia: IEEE, pp. 375-383.

Aral, S., and Weill, P. 2007. " IT Assets, Organizational Capabilities, and Firm Performance: How Resource Allocations and Organizational Differences Explain Performance Variation," *Organization Science* (18:5), pp 763–780.

Baek, W., and Chilimbi, T.M. 2010. "Green: A Framework for Supporting Energy-Conscious Programming Using Controlled Approximation," *ACM SIGPLAN Conference on Programming Language Design and Implementation*, Toronto, Canada, pp. 198-209.

Bharadwaj, A., El Sawy, O.A., Pavlou, P.A., and Venkatraman, N. 2013. "Digital Business Strategy: Toward a Next Generation of Insights," *MIS Quarterly* (37:2), pp 471-482.

Bharadwaj, A.S. 2000. "A Resource-Based Perspective on Information Technology Capability and Firm Performance: An Empirical Investigation," *MIS Quarterly* (24:1), pp 169-196.

Bieser, J.C., and Hilty, L.M. 2018. "Indirect Effects of the Digital Transformation on Environmental Sustainability: Methodological Challenges in Assessing the Greenhouse Gas Abatement Potential of Ict," *International Conference on Information and Communication Technology for Sustainability*, Toronto, Canada: EPiC Series in Computing, pp. 68-81.

Bose, R., and Luo, X. 2011. "Integrative Framework for Assessing Firms' Potential to Undertake Green IT Initiatives Via Virtualization–a Theoretical Perspective," *The Journal of Strategic Information Systems* (20:1), pp 38-54.

Brendel, A.B., Zapadka, P., and Kolbe, L. 2018. "Design Science Research in Green IS-Analyzing the Past to Guide Future Research," in: *European Conference on Information Systems*. Portsmouth, UK: AIS.

Brynjolfsson, E. 2011. *Wired for Innovation: How Information Technology Is Reshaping the Economy*. MIT Press.

Catlin, T., Scanlan, J., and Willmott, P. 2015. "Raising Your Digital Quotient," in: *McKinsey& Company*. McKinsey& Company: McKinsey& Company.

Chan, Y.E., Sabherwal, R., and Thatcher, J.B. 2006. "Antecedents and Outcomes of Strategic IS Alignment: An Empirical Investigation," *IEEE Transactions on Engineering Management* (53:1), pp 27-47.

Cooper, V., and Molla, A. 2017. "Information Systems Absorptive Capacity for Environmentally Driven IS‐Enabled Transformation," *Information Systems Journal* (27:4), pp 379-425.

Corbett, J. 2013. "Designing and Using Carbon Management Systems to Promote Ecologically Responsible Behaviors," *Journal of the Association for Information Systems* (14:7), p 339.

de Medeiros, J.F., Ribeiro, J.L.D., and Cortimiglia, M.N. 2014. "Success Factors for Environmentally Sustainable Product Innovation: A Systematic Literature Review," *Journal of Cleaner Production* (65), 2/15/, pp 76-86.

Degirmenci, K., and Recker, J. 2016. "Boosting Green Behaviors through Information Systems That Enable Environmental Sensemaking," in: *International Conference on Information Systems*. Dublin, Ireland: AIS.

Degirmenci, K., and Recker, J. 2018. "Creating Environmental Sensemaking through Green IS: An Experimental Study on Eco-Nudging Paper Printing Behavior," in: *Americas Conference on Information Systems*. New Orleans, Louisiana, United States: AIS.

Drnevich, P.L., and Croson, D.C. 2013. "Information Technology and Business-Level Strategy: Toward an Integrated Theoretical Perspective," *MIS Quarterly* (37:2), pp 483-509.

Du, S., Bhattacharya, C.B., and Sen, S. 2007. "Reaping Relational Rewards from Corporate Social Responsibility: The Role of Competitive Positioning," *International Journal of Research in Marketing* (24:3), pp 224-241.

El-Kassar, A.-N., and Singh, S.K. 2019. "Green Innovation and Organizational Performance: The Influence of Big Data and the Moderating Role of Management Commitment and Hr Practices," *Technological Forecasting and Social Change* (144), pp 483-498.

Ewing, B., and Baker, E. 2009. "Development of a Green Building Decision Support Tool: A Collaborative Process," *Decision Analysis* (6:3), pp 172-185.

Fielt, E., Bandara, W., Miskon, S., and Gable, G. 2014. "Exploring Shared Services from an IS Perspective: A Literature Review and Research Agenda," *Communications of the Association for Information Systems* (34:1), pp 1001-1040.

Forbes Insight. 2016. "How to Win at Digital Transformation," Jersey City, NJ.

Fuchs, C. 2008. "The Implications of New Information and Communication Technologies for Sustainability," *Environment, Development and Sustainability* (10:3), pp 291-309.

Gerow, J.E., Grover, V., Thatcher, J.B., and Roth, P.L. 2014. "Looking toward the Future of IT-Business Strategic Alignment through the Past: A Meta-Analysis," *MIS Quarterly* (38:4), pp 1059-1085.

Gerow, J.E., Thatcher, J.B., and Grover, V. 2015. "Six Types of IT-Business Strategic Alignment: An Investigation of the Constructs and Their Measurement," *European Journal of Information Systems* (24:5), pp 465-491.

Guster, D., Hemminger, C., and Krzenski, S. 2009. "Using Virtualization to Reduce Data Center Infrastructure and Promote Green Computing," *International Journal of Business Research* (9:6), pp 133-139.

Haffke, I., Kalgovas, B.J., and Benlian, A. 2016. "The Role of the Cio and the Cdo in an Organization's Digital Transformation," in: *International Conference on Information Systems*. Dublin, Ireland: AIS.

Hanelt, A., Busse, S., and Kolbe, L.M. 2016. "Driving Business Transformation toward Sustainability: Exploring the Impact of Supporting IS on the Performance Contribution of Eco-Innovations," *Information Systems Journal* (27:4), pp 463-502.

Hanelt, A., Busse, S., and Kolbe, L.M. 2017. "Driving Business Transformation toward Sustainability: Exploring the Impact of Supporting IS on the Performance Contribution of Eco‐Innovations," *Information Systems Journal* (27:4), pp 463-502.

Hasan, H., Ghose, A., and Spedding, T. 2009. "IS Solution for the Global Environmental Challenge: An Australian Initiative," in: *Americas Conference on Information Systems*. San Francisco, USA: p. 122.

Henderson, J.C., and Venkatraman, N. 1993. "Strategic Alignment: Leveraging Information Technology for Transforming Organizations," *IBM Systems Journal* (32:1), pp 4-16.

Hong, P., Kwon, H.-B., and Roh, J.J. 2009. "Implementation of Strategic Green Orientation in Supply Chain

an Empirical Study of Manufacturing Firms," *European Journal of Innovation Management* (12:4), pp 512-532.

Hong, S.Y., Yang, S.-U., and Rim, H. 2010. "The Influence of Corporate Social Responsibility and Customer–Company Identification on Publics' Dialogic Communication Intentions," *Public Relations Review* (36:2), pp 196-198.

Hu, P.J.-H., Hu, H.-f., Wei, C.-P., and Hsu, P.-F. 2016. "Examining Firms' Green Information Technology Practices: A Hierarchical View of Key Drivers and Their Effects," *Journal of Management Information Systems* (33:4), pp 1149-1179.

Huber, G.P. 1990. "A Theory of the Effects of Advanced Information Technologies on Organizational Design, Intelligence, and Decision Making," *Academy of Management Review* (15:1), pp 47-71.

Jones, N. 2018. "How to Stop Data Centres from Gobbling up the World's Electricity," *Nature* (561:7722), pp 163-167.

Joshi, M.P., Kathuria, R., and Porth, S.J. 2003. "Alignment of Strategic Priorities and Performance: An Integration of Operations and Strategic Management Perspectives," *Journal of Operations Management* (21:3), pp 353-369.

Kappelman, L., McLean, E., Johnson, V., and Gerhart, N. 2014. "The 2014 Sim IT Key Issues and Trends Study," *MIS Quarterly Executive* (13:4), pp 237-263.

Kim, J., Kim, H., and Kwon, H. 2020. "The Impact of Employees' Perceptions of Strategic Alignment on Sustainability: An Empirical Investigation of Korean Firms," *Sustainability* (12:10), pp 1-23.

Kincaid, J. 2009. "How the Times Got Confused About Google and the Tea Kettle," in: *TechCrunch*. Verizon Media.

Kleis, L., Chwelos, P., Ramirez, R.V., and Cockburn, I. 2012. "Information Technology and Intangible Output: The Impact of IT Investment on Innovation Productivity," *Information Systems Research* (23:1), pp 42-59.

Kumar, V., Aksoy, L., Donkers, B., Venkatesan, R., Wiesel, T., and Tillmanns, S. 2010. "Undervalued or Overvalued Customers: Capturing Total Customer Engagement Value," *Journal of Service Research* (13:3), pp 297-310.

Lampikoski, T., Westerlund, M., Rajala, R., and Möller, K. 2014. "Green Innovation Games," *California Management Review* (57:1), pp 88-116.

Li, L., Su, F., Zhang, W., and Mao, J.Y. 2018. "Digital Transformation by Sme Entrepreneurs: A Capability Perspective," *Information Systems Journal* (28:6), pp 1129-1157.

Lieb, K.J., and Lieb, R.C. 2010. "Environmental Sustainability in the Third‐Party Logistics (3pl) Industry," *International Journal of Physical Distribution & Logistics Management* (40:7), pp 524-533.

Loeser, F., Recker, J., Brocke, J.v., Molla, A., and Zarnekow, R. 2017. "How IT Executives Create Organizational Benefits by Translating Environmental Strategies into Green IS Initiatives," *Information Systems Journal*).

Lokuge, S., and Sedera, D. 2014a. "Deriving Information Systems Innovation Execution Mechanisms," *Australasian Conference on Information Systems*, Auckland, New Zealand: AIS.

Lokuge, S., and Sedera, D. 2014b. "Enterprise Systems Lifecycle-Wide Innovation Readiness," *Pacific Asia Conference on Information Systems*, Chengdu, China: AIS.

Lokuge, S., and Sedera, D. 2016. "Is Your IT Eco-System Ready to Facilitate Organizational Innovation? Deriving an IT Eco-System Readiness Measurement Model," *International Conference on Information Systems*, Dublin, Ireland: AIS.

Lokuge, S., and Sedera, D. 2018. "The Role of Enterprise Systems in Fostering Innovation in Contemporary Firms," *Journal of Information Technology Theory and Application (JITTA)* (19:2), pp 7-30.

Lokuge, S., and Sedera, D. 2019. "Attaining Business Alignment in Information Technology Innovations Led by Line-of-Business Managers," in: *Australasian Conference on Information Systems*. Perth, Australia.

Lokuge, S., and Sedera, D. 2020. "Fifty Shades of Digital Innovation: How Firms Innovate with Digital Technologies," *Pacific Asia Conference on Information Systems*, Dubai, UAE: AIS, p. 91.

Lokuge, S., Sedera, D., Ariyachandra, T., Kumar, S., and Ravi, V. 2020. "The Next Wave of Crm Innovation: Implications for Research, Teaching, and Practice," *Communications of the Association for Information Systems* (46:1), pp 560-583.

Lokuge, S., Sedera, D., Grover, V., and Xu, D. 2019. "Organizational Readiness for Digital Innovation: Development and Empirical Calibration of a Construct," *Information & Management* (56:3), pp 445-461.

Lokuge, S., Sedera, D., and Perera, M. 2018. "The Clash of the Leaders: The Intermix of Leadership Styles for Resource Bundling," *Pacific Asia Conference on Information Systems*, Yokohama, Japan: AIS.

Lopez, J. 2014. "Digital Business Is Everyone's Business," in: *Forbes*.

Luo, X., and Du, S. 2012. "Good" Companies Launch More New Products," *Harvard Business Review* (90:4), p 28.

Majchrzak, A., Markus, L.M., and Wareham, J. 2016. "Designing for Digital Transformation: Lessons for Information Systems Research from the Study of Ict and Societal Challenges," *MIS Quarterly* (40:2), pp 267-277.

McKinnon, A. 2010a. "Environmental Sustainability," in: *Green Logistics: Improving the Environmental Sustainability of Logistics. London,* A. McKinnon, S. Cullinane, M. Browne and A. Whiteing (eds.). London, UK: Kogan Page Limited.

McKinnon, A. 2010b. "The Role of Government in Promoting Green Logistics," in: *Green Logistics: Improving the Environmental Sustainability of Logistics. London,* A. McKinnon, S. Cullinane, M. Browne and A. Whiteing (eds.). London, UK: Kogan Page Limited.

McLaren, T.S., Head, M.M., Yuan, Y., and Chan, Y.E. 2011. "A Multilevel Model for Measuring Fit between a Firm's Competitive Strategies and Information Systems Capabilities," *MIS Quarterly* (35:4), pp 909-929.

Melville, N.P. 2010. "Information Systems Innovation for Environmental Sustainability," *MIS Quarterly* (34:1), pp 1-21.

Melville, N.P., and Zik, O. 2016. "Energy Points: A New Approach to Optimizing Strategic Resources by Leveraging Big Data," *Hawaii International Conference on System Sciences*, Kauai, USA: IEEE, pp. 1030-1039.

Miguel, R.S. 2009. "Harvard Physicist Sets Record Straight on Internet Carbon Study," in: *TechNewsWorld*. ECT News Network.

Molla, A., and Abareshi, A. 2012. "Organizational Green Motivations for Information Technology: Empirical Study," *Journal of Computer Information Systems* (52:3), 2012/03/01, pp 92-102.

Molla, A., Cooper, V., and Pittayachawan, S. 2011. "The Green IT Readiness (G-Readiness) of Organizations: An Exploratory Analysis of a Construct and Instrument," *Communications of the Association for Information Systems* (29:1), p 4.

Nishant, R., Teo, T., Goh, M., and Krishnan, S. 2012. "Does Environmental Performance Affect Organizational Performance? Evidence from Green IT Organizations," *International Conference on Information Systems*, Orlando: AIS.

Nylén, D., and Holmström, J. 2015. "Digital Innovation Strategy: A Framework for Diagnosing and Improving Digital Product and Service Innovation," *Business Horizons* (58:1), pp 57-67.

Olesen, K., and Myers, M.D. 1999. "Trying to Improve Communication and Collaboration with Information Technology: An Action Research Project Which Failed," *Information Technology & People* (12:4), pp 317-332.

Queiroz, M., Coltman, T., Tallon, P., Sharma, R., and Reynolds, P. 2018. "The Complementarity of Corporate IT Alignment and Business Unit IT Alignment: An Analysis of Their Joint Effects on Business Unit Performance," *Hawaii International Conference on System Sciences*, Hawaii, USA: AIS.

Rush, D., Melville, N., Ramirez, R., and Kobelsky, K. 2015. "Enterprise Information Systems Capability and Ghg Pollution Emissions Reductions," *International Conference on Information Systems*, Fort Worth, Texas: AIS.

Sabherwal, R., and Chan, Y. 2001. "Alignment between Business and IT Strategies: A Study of Prospectors, Analyzers and Defenders," *Information Systems Research* (12:1), pp 11–33.

Schiederig, T., Tietze, F., and Herstatt, C. 2012. "Green Innovation in Technology and Innovation Management–an Exploratory Literature Review," *R&D Management* (42:2), pp 180-192.

Sedera, D., and Lokuge, S. 2017. "The Role of Enterprise Systems in Innovation in the Contemporary Organization," in: *The Routledge Companion to Management Information Systems,* R.G. Galliers and M.-K. Stein (eds.). Abingdon, United Kingdom: The Routledge p. 608.

Sedera, D., and Lokuge, S. 2019. "Do We Put All Eggs in One Basket? A Polynomial Regression Study of Digital Technology Configuration Strategies," in: *International Conference on Information Systems.* Munich, Germany: AIS.

Sedera, D., Lokuge, S., Grover, V., Sarker, S., and Sarker, S. 2016. "Innovating with Enterprise Systems and Digital Platforms: A Contingent Resource-Based Theory View," *Information & Management* (53:3), pp 366–379.

Sedera, D., Lokuge, S., Tushi, B., and Tan, F. 2017. "Multi-Disciplinary Green IT Archival Analysis: A Pathway for Future Studies," *Communications of the Association for Information Systems* (41:1), pp 674-733.

Seidel, S., Chandra Kruse, L., Székely, N., Gau, M., and Stieger, D. 2018. "Design Principles for Sensemaking Support Systems in Environmental Sustainability Transformations," *European Journal of Information Systems* (27:2), pp 221-247.

Sendall, P., Shannon, L.-J.Y., Peslak, A., and Saulnier, B. 2011. "The Greening of the Information Systems Curriculum," *Information Systems Education Journal* (9:5), pp 27-45.

Singh, R., Mathiassen, L., Stachura, M.E., and Astapova, E.V. 2011. "Dynamic Capabilities in Home Health: IT-Enabled Transformation of Post-Acute Care," *Journal of the Association for Information Systems* (12:2), pp 163-188.

Suchman, M.C. 1995. "Managing Legitimacy: Strategic and Institutional Approaches," *Academy of Management Review* (20:3), pp 571-610.

Sui, D.Z., and Rejeski, D.W. 2002. "Environmental Impacts of the Emerging Digital Economy: The E-for-Environment E-Commerce?," *Environmental Management* (29:2), pp 155-163.

Teece, D.J., Pisano, G., and Shuen, A. 1997. "Dynamic Capabilities and Strategic Management," *Strategic Management Journal* (18:7), pp 509-533.

The World Bank. 2019. "Climate Change." *Understanding Poverty*, 2019, from https://www.worldbank.org/en/topic/climatechange/overview

Tsai, W.-H., Tsaur, T.-S., Chou, Y.-W., Liu, J.-Y., Hsu, J.-L., and Hsieh, C.-L. 2015. "Integrating the Activity-Based Costing System and Life-Cycle Assessment into Green Decision-Making," *International Journal of Production Research* (53:2), pp 451-465.

Turnhout, E., Dewulf, A., and Hulme, M. 2016. "What Does Policy-Relevant Global Environmental Knowledge Do? The Cases of Climate and Biodiversity," *Current Opinion in Environmental Sustainability* (18), pp 65-72.

Tushi, B., Sedera, D., and Recker, J. 2014. "Green IT Segment Analysis: An Academic Literature Review," in: *Americas Conference on Information Systems*. Georgia, USA: AIS.

Vahabzadeh, A.H., Asiaei, A., and Zailani, S. 2015. "Green Decision-Making Model in Reverse Logistics Using Fuzzy-Vikor Method," *Resources, Conservation and Recycling* (103), pp 125-138.

Venable, J.R., Pries‐Heje, J., Bunker, D., Russo, N.L., Venable, J.R., Bunker, D., and Russo, N.L. 2011. "Design and Diffusion of Systems for Human Benefit," *Information Technology & People* (24:3), pp 208-216.

Vial, G. 2019. "Understanding Digital Transformation: A Review and a Research Agenda," *The Journal of Strategic Information Systems* (28:2), pp 118-144.

Wade, M., and Hulland, J. 2004. "Review: The Resource-Based View and Information Systems Research: Review, Extension, and Suggestions for Future Research," *MIS Quarterly* (28:1), pp 107-142.

Walther, S., Sarker, S., Urbach, N., Sedera, D., Eymann, T., and Otto, B. 2015. "Exploring Organizational Level Continuance of Cloud-Based Enterprise Systems," in: *European Conference on Information Systems*. Münster, Germany: AIS.

Walther, S., Sedera, D., Urbach, N., Eymann, T., Otto, B., and Sarker, S. 2018. "Should We Stay, or Should We Go? Analyzing Continuance of Cloud Enterprise Systems," *Journal of Information Technology Theory and Application* (19:2), pp 57-88.

Wang, X., Brooks, S., and Sarker, S. 2015. "Understanding Green IS Initiatives: A Multi-Theoretical Framework," *Communications of the Association for Information Systems* (37:32).

Watson, R.T., Boudreau, M.-C., and Chen, A.J. 2010. "Information Systems and Environmentally Sustainable Development: Energy Informatics and New Directions for the IS Community," *MIS Quarterly* (34:1), pp 23-38.

Wessel, L., Baiyere, A., Ologeanu-Taddei, R., Cha, J., and Jensen, T. 2020. "Unpacking the Difference between Digital Transformation and IT-Enabled Organizational Transformation," *Journal of Association of Information Systems*).

World Commission on Environment Development. 1987. "Report of the World Commission on Environment and Development: Our Common Future," World Commission on Environment and Development, Brussels.

World Economic Forum. 2016. "Global Risks Report 2016," World Economic Forum, https://www.weforum.org/.

Youmans, W.L., and York, J.C. 2012. "Social Media and the Activist Toolkit: User Agreements, Corporate Interests, and the Information Infrastructure of Modern Social Movements," *Journal of Communication* (62:2), pp 315-329.

# About the Authors

**Sachithra Lokuge** is a Lecturer in Information Systems at the College of Business and Law at the RMIT University, Australia. Her research mainly focuses on digital technologies, innovation, digital entrepreneurship and social media use. Her work has been published in journals such as the Information & Management, Information Technology and People, Communications of the Association for Information Systems, Journal of Information Technology Theory and Application (JITTA), and in the proceedings of conferences such as the International Conference on Information Systems and Academy of Management.

**Darshana Sedera** is a Professor of Information Systems and the Director of the Digital Enterprise Lab at the Southern Cross University in Australia. He is currently the Deputy Dean of the School of Business and Tourism. He has published over 200 publications in major refereed journals and conferences. His publications have appeared in Journal of the Association for Information Systems, Journal of the Strategic Information Systems, Information & Management, Information Technology & People, and Communications of the Association for Information Systems.

**Vanessa Cooper** is a Professor in Information Systems at RMIT University, Melbourne, Australia and leader of the Digital Work and Society research program in the Centre for People, Organization and Work. Her research in information systems centers on the application of knowledge management and organizational learning theory in contexts such as environmental sustainability, emergency management and the future of work. Her research has been published in leading academic journals and won multiple awards, including best paper awards at international conferences and from the Australian Computer Society.

**Frada Burstein** is a Professor of Information Systems at the Department of Human-Centered Computing, Faculty of Information Technology, Monash University. She was the Deputy Director of the Centre for Organizational and Social Informatics and led the Community Health and Wellbeing Informatics theme for the research flagship "IT for Resilient Communities". Her professional academic career spans more than twenty years and includes extensive research in the areas of decision support systems, knowledge management, and more recently in digital transformation of healthcare. In 2013 Professor Frada Burstein was named the ICT Educator of the Year for her contribution to information and communications technology (ICT) education. Prof Burstein is a Fellow of Australian Computer Society and a Distinguished Member of the Association for Information Systems. Her research appears in such journals as Decision Support Systems, Journal of the American Society for Information Science and Technology, Journal of IT, Journal of Knowledge Management, Knowledge Management Research and Practice and others.